\documentclass[prl,aps,showpacs,twocolumn]{revtex4}
\usepackage{graphicx}

\begin{document}

\title{Manifestation of finite temperature size effects in nanogranular magnetic graphite}

\author{S. Sergeenkov$^{1}$,  N.S. Souza$^{1}$, C. Speglich$^{1}$, V.A.G. Rivera$^{1}$,
C.A. Cardoso$^{1}$,  H. Pardo$^{2}$, A.W. Mombr\'{u}$^{2}$ and
F.M. Ara\'{u}jo-Moreira$^{1}$}

\affiliation{$^{1}$Materials and Devices Group, Department of
Physics and Physical Engineering, Universidade Federal
de S\~ao Carlos, S\~ao Carlos, SP, 13565-905 Brazil\\
$^{2}$Crystallography, Solid State and Materials Laboratory
(Cryssmat-Lab), DEQUIFIM, Facultad de Qu\'{i}mica, Universidad de
la Rep\'{u}blica, P.O. Box 1157, CP 11800, Montevideo, Uruguay}

\date{\today}

\begin{abstract}
In addition to the double phase transition (with the Curie
temperatures $T_C=300K$ and $T_{Ct}=144K$), a low-temperature
anomaly in the dependence of the magnetization is observed in the
bulk magnetic graphite (with an average granular size of $L\simeq
10nm$), which is attributed to manifestation of the size effects
below the quantum temperature $T_L\propto \hbar^2/L^2$ and is well
fitted by the periodic function $M_L(T)\propto \sin[M(T)\Lambda
(T)/L]$ with $M(T)$ being the bulk magnetization and $\Lambda
(T)\propto \hbar/\sqrt{T}$ the thermal de Broglie wavelength. The
best fits of the high-temperature data (using the mean-field
Curie-Weiss and Bloch expressions) produced reasonable estimates
for the model parameters, such as defects mediated effective spin
exchange energy $J\simeq 12meV$ (which defines the intragranular
Curie temperature $T_C$) and proximity mediated interactions
between neighboring grains (through potential barriers $U$ created
by thin layers of non-magnetic graphite) with energy
$J_t=\exp(-d/\xi )J\simeq 5.8meV$ (which defines the intergranular
Curie temperature $T_{Ct}$) with $d\simeq 1.5nm$ and $\xi \propto
\hbar/\sqrt{U}\simeq 2nm$ being the intergranular distance and
characteristic length, respectively.
\end{abstract}

\pacs{75.50.Dd, 78.70.2g, 81.05.Uw}

\maketitle

Recent advances in developing nanographitic systems re-kindled
interest in their nontrivial magnetic and transport properties
important for numerous applications (for recent reviews, see,
e.g.,~\cite{1,2,3} and further references therein). Special
attention has been paid to the properties mediated by various
defect structures (including pores, edges of the planes,
chemically induced vacancies, dislocations, tracks produced by
particle irradiation, etc) believed to be responsible for
room-temperature ferromagnetism based on (super) exchange between
localized spins at defect sites. The existence of sufficiently
robust ferromagnetic (FM) like magnetization loops has been
successfully proved in highly-oriented pyrolytic graphite
(HOPG)~\cite{4}, proton-irradiated graphite~\cite{5},
nanographite~\cite{6}, graphite containing topographic
defects~\cite{7}, negative curvature Schwarzite-like carbon
nanofoams~\cite{8}, fullerene-related carbons~\cite{9},
microporous carbon~\cite{10}, and carbon nanohorns~\cite{11}.
Recently, some interesting results have been reported~\cite{12}
regarding unusual magnetic properties of $Ag$ nanoparticles
encapsulated in carbon nanospheres (with $\simeq 10nm$ diameter)
interconnected in necklace-like structures which have a tremendous
potential for applications in electronics, biotechnology and
medicine.

In this paper, we report our latest results on the temperature
dependence of the magnetization in bulk room-temperature magnetic
graphite (MG) with an average grain size of $L\simeq 10nm$.
Several interesting features have been observed in our MG samples,
including (i) the double transition with the Curie temperatures
$T_C=300K$ and $T_{Ct}=144K$ attributed, respectively, to the
manifestation of the intragranular $M_p(T)$ and intergranular
$M_t(T)$ contributions to bulk magnetization $M(T)$, and (ii) a
low-temperature anomaly in the dependence of $M(T)$, attributed to
manifestation of the finite temperature size effects below the
quantum temperature $T_L\propto \hbar^2/L^2$.

Our MG samples were produced by a vapor phase redox controlled
reaction in closed nitrogen atmosphere with addition of copper
oxide using synthetic graphite powder (more details regarding the
patented chemical route for synthesis of the discussed here
magnetic graphite can be found elsewhere~\cite{13,14}). To avoid
presence of any kind of FM impurity, we have carefully determined
the chemical purity of the samples with atomic absorption
spectroscopy (AAS) using a Shimadzu $AA6800$ spectrometer and
checked these results with X-ray fluorescence analysis (XRF) and
energy dispersive spectroscopy (EDS), comparing the results
obtained for the pristine (non-magnetic) and the modified
(magnetic) graphite. The structure of MG samples has been verified
by Raman spectroscopy, X-ray diffraction (XRD) and scanning
electron microscopy (SEM). These studies were performed using
Seifert Scintag PAD-II powder diffractometer, with $CuK\alpha$
radiation ($\lambda = 1.5418$\AA) and  Jeol JSM $5900LV$
microscope, respectively. In addition to the broader (as compared
with the pristine non-magnetic graphite) peak at $1580 cm^{-1}$,
corresponding to chemically modified magnetic graphite, our micro
Raman analysis shows the appearance of a new peak at $1350
cm^{-1}$ (known as the "disordered" $D$ band) in the MG sample. In
turn, the XRD profiles revealed that the peaks of the magnetic
graphite for $(002)$ and $(004)$ reflections are wider and
asymmetric, with a visible compression of $c$-axis (due to
chemically induced defects in the MG structure) that facilitates
bringing the graphene layers closer to each other (thus further
enhancing FM properties of the sample). To verify the correlation
between the microstructural features (topography) and the presence
of magnetic regions in MG sample, we also used the atomic force
(AFM)  and magnetic force (MFM) microscopy. The comparison of the
obtained AFM and MFM $3D$ images (along with the corresponding SEM
images) revealed that our MG sample is a rather dense
agglomeration of spherical particles (with diameters of $L\simeq
10nm$) coated by thin layers of non-reacted pristine
(non-magnetic) graphite (with thickness of $d\simeq 1-2nm$),
producing both intra- and intergranular magnetic response.

The magnetization measurements were done using a MPMS-$5T$ Quantum
Design magnetometer. Both zero-field cooled (ZFC) and field-cooled
(FC) $M-T$ cycles were measured. From the $M-H$ hysteresis loop
for MG sample (with mass $0.4mg$) taken at $T=295K$ and after
subtracting diamagnetic background (equivalent to $1.2\times
10^{-3}\mu _B$ per carbon atom), we deduced $M_s=0.25 emu/g$,
$M_r=0.04emu/g$ and $H_C=350Oe$ for the room-temperature values of
saturation magnetization, remnant magnetization and coercive
field, respectively. The temperature behavior of the normalized
ZFC magnetization $M(T)/M(T_p)$ in our MG sample (taken at
$H=1kOe$) is shown in Fig.1 after subtracting the diamagnetic and
paramagnetic contributions ($T_p=0.16T_C=48K$ is the temperature
where $M(T)$ has a maximum with the absolute value of
$M(T_p)=0.12emu/g$). First of all, notice that there are two
distinctive regions, below and above the peak temperature $T_p$.
Namely, below $T_p$ there is a well-defined low-temperature
minimum (around $T_{m}=0.05T_C= 15K$), while for $T>T_p$ we have a
crossover region (near $T_0=0.38T_C=114K$) indicating the presence
of a double phase transition in our sample. More precisely, in
addition to the phase with the Curie temperature $T_C=300K$, there
is a second transition with $T_{Ct}=0.48T_C=144K$.

\begin{figure}
\centerline{\includegraphics[width=8.0cm]{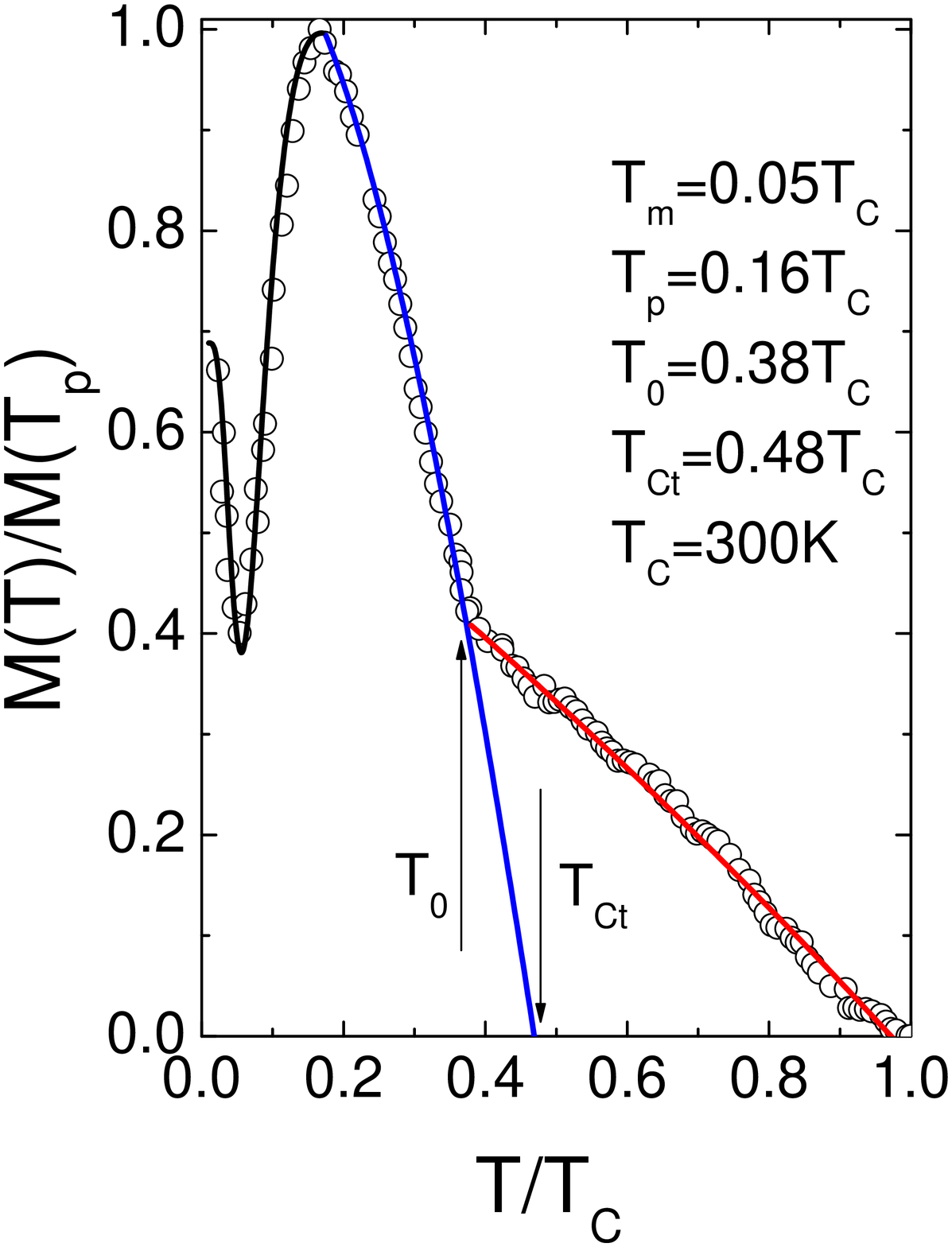}}\vspace{0.5cm}
\caption{The temperature dependence of the normalized
magnetization $M(T)$ of magnetic graphite. The solid lines are the
best fits according to Eqs.(1)-(4).}
\end{figure}

Let us begin our discussion with the high-temperature region
(above $T_p$). By attributing $T_C$ and $T_{Ct}$ to the
manifestation of the intrinsic $M_p$ and extrinsic (intergranular)
$M_t$ contributions to the observed magnetization $M(T)$,
respectively, we were able to successfully fit our data using the
following expressions:
\begin{equation}
M(T)=M_p(T)+M_t(T)
\end{equation}
with
\begin{equation}
M_p(T)=M_{sp}\tanh{\sqrt{\left(\frac{T_C}{T}\right)^2-1}}
+M_{mp}\left[1-\left(\frac{T}{T_C}\right)^{3/2} \right]
\end{equation}
and
\begin{equation}
M_t(T)=M_{st}\tanh{\sqrt{\left(\frac{T_{Ct}}{T}\right)^2-1}}
+M_{mt}\left[1-\left(\frac{T}{T_{Ct}}\right)^{3/2}\right]
\end{equation}
The first terms in the rhs of  Eqs.(2) and (3) present  analytical
(approximate) solution of the Curie-Weiss mean-field equation for
spontaneous magnetization valid for all temperatures (see,
e.g.,~\cite{15,16}), while the second terms account for the Bloch
(magnon) contributions~\cite{17}. The solid lines in Fig.1 present
the best fits for high-temperature region ($T\ge T_p$) according
to Eqs.(1)-(3) with the following set of parameters (in terms of
the experimental value of the total magnetization
$M(T_p)=0.12emu/g$): $M_{sp}=0.59M(T_p)$, $M_{mp}=0.11M(T_p)$,
$T_C=300K$, $M_{st}=0.29M(T_p)$, $M_{mt}=0.05M(T_p)$, and
$T_{Ct}=144K$. Notice that the above estimates suggest quite a
significant contribution from the intergranular interactions
($M_t\simeq 0.5M_p$).

To better understand the origin of the model parameters, recall
that, according to recent theoretical analysis~\cite{5,12,18}, the
room-temperature FM in graphite is most likely due to
superexchange mediated by the two different sites in the graphite
lattice leading to a FM coupling between localized spins $S$ at
the defect sites with an effective exchange energy $J$, related to
the intragranular Curie temperature $T_C=S(S+1)zJ/3k_B$ (here $z$
is the number of nearest neighbors). As is well-known, graphite is
made of two-dimensional layers in which each carbon is covalently
bonded to three other carbons. Atoms in other layers are much
further away and are not nearest neighbors, so the coordination
number of a carbon atom in graphite is $z=3$. Using $S=1/2$ and
the experimentally found $T_C =300K$, we obtain $J\simeq 12meV$
for a reasonable estimate~\cite{5,18} of the defects mediated spin
exchange coupling energy (responsible for the intragranular
contribution $M_p(T)$). Besides, within this scenario, the deduced
from our $M-H$ hysteresis loops value of the room-temperature
saturation magnetization $M_s=0.25emu/g$ corresponds to defect
concentration of $\simeq 600 ppm$, which is within the range
reported for nanographite-based carbon materials~\cite{19} and is
high enough to account for the observed strong FM like response.

At the same time, given the above-discussed chemically modified
nanogranular structure in our sample, it is quite reasonable to
assume that the second transition with $T_{Ct}=144K$ (responsible
for the intergranular contribution $M_t(T)$) is related to the
proximity mediated tunneling of the delocalized spins between
neighboring grains (through potential barriers $U$ created by thin
layers of pristine non-magnetic graphite) with the probability
$J_t=\exp(-d/\xi )J$. Here, $d$ is the distance between adjacent
particles and $\xi= \hbar /\sqrt{2m^{*}U}$ is a characteristic
length with $m^{*}$ being the effective mass. According to this
scenario, the intergranular Curie temperature $T_{Ct}$ is related
to its intragranular counterpart as $T_{Ct}=\exp(-d/\xi )T_C$.
Furthermore, by correlating the crossover temperature
$T_0=0.38T_C$ with the value of the intergranular barrier $U\simeq
k_BT_0$, we obtain $U\simeq 8meV$ for its estimate (assuming free
electron mass for $m^{*}$) which, in turn, brings about $\xi
\simeq 2nm$ for an estimate of the characteristic length.
Moreover, using the found values of the Curie temperatures
($T_C=300K$ and $T_{Ct}=144K$), we obtain $d\simeq 1.5nm$ as a
reasonable estimate for an average thickness of non-magnetic
graphite layer between magnetic particles in our MG sample. It is
also interesting to observe that, given the above obtained value
for the tunneling exponent $\exp(-d/\xi )\simeq 0.48$, the
relations between the intra- and intergranular fitting parameters,
$M_{st}=0.49M_{sp}$ and $M_{mt}=0.47M_{ms}$, are in good agreement
with the proximity mediated scenario, assuming $M_{st}=\exp(-d/\xi
)M_{sp}$ and $M_{mt}=\exp(-d/\xi )M_{ms}$ for the Curie-Weiss and
Bloch magnetizations.

\begin{figure}
\centerline{\includegraphics[width=8.0cm]{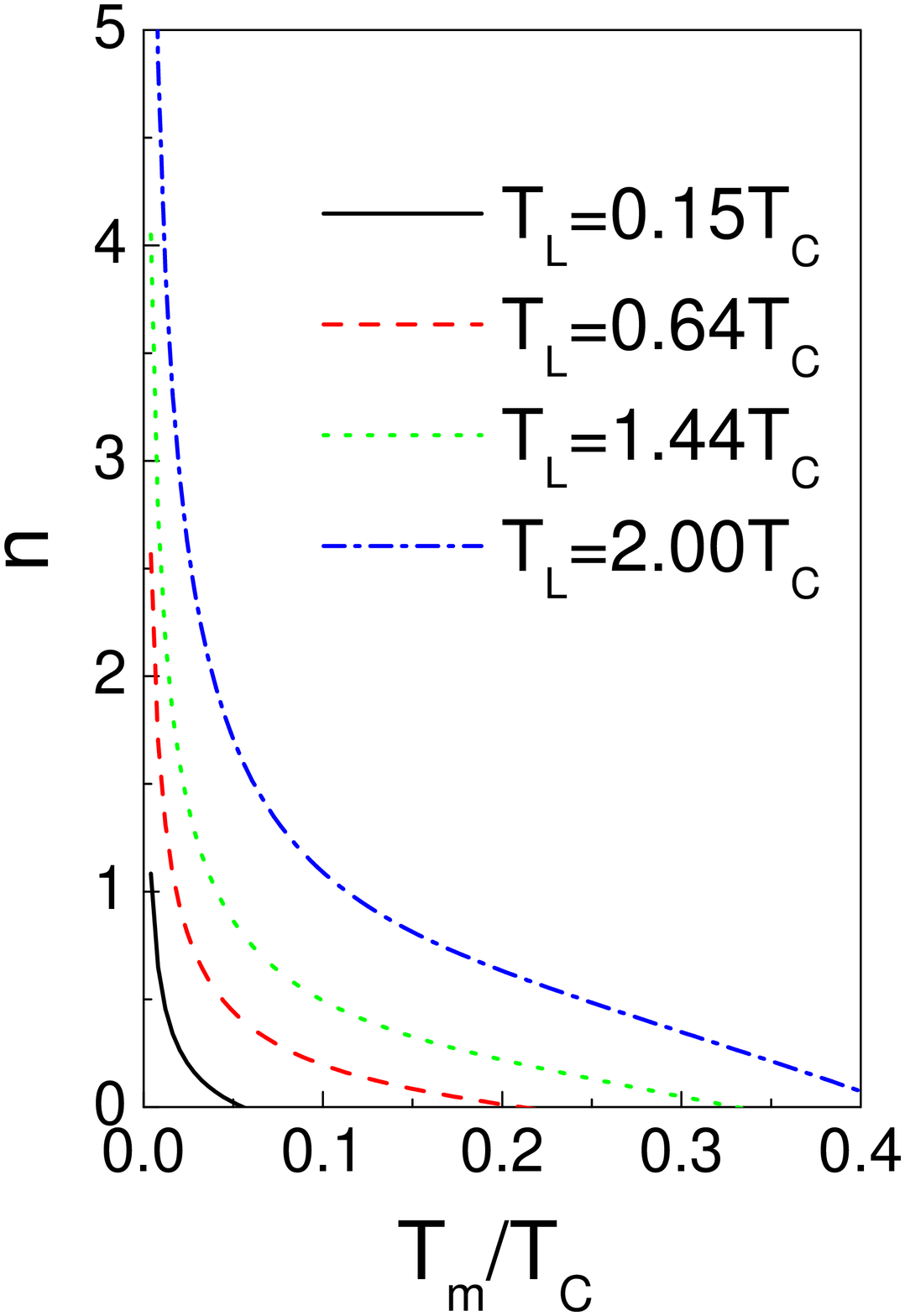}}\vspace{0.5cm}
\caption{The dependence of the oscillations minima $n$ on reduced
temperature $T_m/T_C$ for different values of the particle size
$L$ related quantum temperature $T_L$, according to Eqs.(1)-(4).}
\end{figure}

Let us turn now to the low-temperature region ($T<T_p$) and
discuss the origin of the observed minimum of magnetization near
$T_m=0.05T_C$. We will show that this anomaly can be attributed to
the quantum size effect. Recall that the finite temperature
quantum effects manifest themselves for the size of the system
$L<\Lambda (T)$ (where $\Lambda (T)=\sqrt{2\pi \hbar^2/m^{*}k_BT}$
is the thermal de Broglie wavelength) or, alternatively, for
temperatures $T<T_L$ (where $T_L=2\pi \hbar^2/m^{*}k_BL^2$ is the
quantum temperature). Using $L\simeq 10nm$ for an average size of
the single particle in our samples (and assuming free electron
mass for $m^{*}$), we get $T_L=0.15T_C= 45K$ for the onset
temperature below which the manifestation of quantum size effects
is expected (notice that $T_L$ is very close to the peak
temperature $T_p=0.16T_C$). To fit the low-temperature
experimental data, we assume the following normalized (to the peak
temperature $T_p$) periodic dependence of the finite-size
magnetization:
\begin{equation}
\frac{M_L(T)}{M_L(T_p)}=\left[\frac{L}{\Lambda(T)}\right]\sin\left\{\left[
\frac{M(T)}{M(T_p)}\right]\left[\frac{\Lambda(T)}{L}\right]\right\}
\end{equation}
where $M(T)$ is the above-considered total bulk magnetization
(thus we assume that quantum effects will influence both intra-
and intergrain properties). It can be easily verified that Eq.(4)
reduces to $M(T)$ when the quantum effects become negligible. More
precisely, $M(T)/M(T_p)=\lim_{L\gg \Lambda(T)}[M_L(T)/M_L(T_p)]$.
The best fit of the low-temperature region ($T<T_p$) using
Eqs.(1)-(4) is shown by thick solid line in Fig.1. Notice also
that, for a given temperature, the above periodic function
$M_L(T)$ has minima at $T=T_m$ where $T_m$ is the solution of the
following equation, $M(T_m)\Lambda (T_m)=\pi (n+1)M(T_p)L$ with
$n=0,1,2,...$ being the number of the oscillation minima. Using
the explicit temperature dependencies of the total bulk
magnetization $M(T)$ (given by Eqs.(1)-(3)) and the previously
defined thermal de Broglie wavelength $\Lambda(T)$, in Fig.2 we
depict the solution of the above equation as the dependence of the
quantization minima $n$ on reduced temperature $T_m/T_C$ for
different values of the particle size $L$ (in terms of the quantum
temperature $T_L\propto \hbar^2/L^2$). As it is clearly seen in
this picture, the smaller the particle size (hence, the larger the
quantization temperature $T_L$), the more finite size related
oscillations (minima) should be observed in the temperature
dependence of the magnetization $M_L(T)$. For example, in our
particular case (with $L=10nm$ and $T_L=0.15T_C$) only "ground
state" minimum (corresponding to $n=0$) is expected to be visible
at non-zero temperature $T_m=0.05T_C=15K$, in agreement with the
observations (see Fig.1). And finally, it should be mentioned that
a similar magnetization peak around $50K$ has been also observed
in carbon nanohorns~\cite{20} and carbon nanosphere
powder~\cite{12}, where its origin was attributed to adsorbed
oxygen and first-order spin reorientation transition,
respectively.

In summary, some interesting experimental results  related to low-
and high-temperature features of the zero-field-cooled
magnetization in bulk magnetic graphite have been presented and
discussed. The proposed theoretical interpretation for
intragranular and intergranular contributions was based,
respectively,  on superexchange interaction between defects
induced localized spins in a single grain and proximity mediated
interaction between grains (through the barriers created by thin
layers of non-magnetic graphite).

This work has been financially supported by the Brazilian agencies
CNPq, CAPES and FAPESP.

\end{document}